\documentclass[a4paper,10pt]{article}
\usepackage[dvips]{graphicx}
\usepackage{amssymb,amsmath}
\numberwithin{equation}{section}

\begin{document}

\title{Accelerating universe from F(T)   gravity}
\author{Ratbay Myrzakulov\footnote{Email: rmyrzakulov@gmail.com, rmyrzakulov@csufresno.edu} \\ \textit{Eurasian International Center for Theoretical Physics,} \\ \textit{ Eurasian National University, Astana 010008, Kazakhstan}
\\ and \\ \textit{Department of Physics, CSU Fresno, Fresno, CA 93740 USA}}

\date{}

\maketitle

\begin{abstract}
It is shown that the acceleration of the universe can be understood by considering a $F(T)$ gravity models. For these $F(T)$ gravity models,  a variant of the accelerating cosmology reconstruction program is developed. Some explicit examples of $F(T)$ are reconstructed from the background FRW expansion history. 
\end{abstract}
\vspace{2cm} \sloppy


\section{Introduction}
Recent observational data indicate that our universe is accelerating. This acceleration is explained in terms of the so-called dark energy (DE), which may be explained in modified gravity models (for a general review see Ref.\cite{N1}). DE could also result from a cosmological constant, from an ideal fluid with a different form of equation of state (EoS) and negative pressure, a scalar field with quintessence-like or phantom-like behavior (see \cite{Eli1} and references therein), etc. The choice of possibilities reflects the indisputable fact that the true nature and origin of DE has not been convincingly explained yet. It is not even clear what type of DE is more seemingly to explain the current epoch of the universe. Observational data point toward some kind of DE with an EoS parameter which is very close to $w=-1$, maybe even less than $-1$ (the so-called phantom case). A quite appealing possibility is the already mentioned modification of General Relativity (GR). Modifications of the Hilbert-Einstein action by introducing different functions of the Ricci scalar $R$ have been systematically explored, the so-called $F(R)$ gravity models, which reconstruction has been developed in Refs.~\cite{N7}-\cite{DSG}. 
As is known, $F(R)$  gravity can be written in terms of a scalar field - quintessence or phantom like - by redefining the function $F(R)$  with the use of a scalar field, and then performing a conformal transformation. It has been shown that, in general, for any given $F(R)$ the corresponding scalar-tensor theory can, in principle, be obtained, although
the solution is going to be very different from one case to another. Also, attention has been paid to the reconstruction of $F(R)$ gravity from a given scalar-tensor theory. It is known, too, that the phantom case in scalar-tensor theory does not exist, in general, when starting from $F(R)$ gravity. In fact, the conformal transformation becomes complex
when the phantom barrier is crossed, and therefore the resulting $F(R)$ function becomes complex. These situations where addressed in \cite{DSG} in detail, where to avoid this hindrance, a dark fluid was used in order to produce the phantom behavior in such a way that the $F(R)$ function reconstructed from the scalar-tensor theory continues to be real.

On the other hand, it has also been suggested in the literature (see e.g.  \cite{N2} -- \cite{N9}) to consider modified Gauss-Bonnet gravity, that is, a function of the GB invariant. Different cosmological properties of modified gravity models of this kind have been studied in \cite{N2}-\cite{DGOG}. Both possibilities have, in principle, the capability to explain the accelerated expansion of the Universe and even the primordial inflationary phase (see \cite{N7} and \cite{DSG}), with no need to introduce a new form of energy. 
So, a quite appealing possibility for the gravitational origin of the DE is the modification of GR. Actually, there is no compelling reason why standard  GR should be trusted at large cosmological scales.

 Unfortunately, many theoretical models that anyhow describing the accelerated expansion of the universe  and which appears to fit all currently available observations (Supernovae Ia, CMBR anisotropies, Large Scale Structure formation, baryon oscillations and weak lensing)  are affected by significant fine-tuning problems related to the vacuum energy scale and therefore it is important to investigate alternatives to this description of the Universe.  There are exist several other approaches to the theoretical description of the accelerated expansion of the universe. As we mentioned above, one of these is a modified gravity theories.  Studies of the physics of these theories is however hampered by the complexity of the field equations, making it difficult to obtain both exact and numerical solutions which can be compared with observations. These problems can be reduced somewhat by using the theory of dynamical systems \cite{WE}, which provides a relatively simple method for obtaining exact solutions and a description of the global dynamics of these models for a given $F(R)$ theory. Another approach to modified gravity theories is the reconstruction method (see e.g. for review \cite{N1} and references therein).
 Another interesting sort of modified theories is so-called $F(T)$ - gravity ($T$ is torsion) \cite{FF1}-\cite{FF2}. Recently,   it is shown (\cite{FF1}-\cite{Boehmer}) that such $F(T)$-gravity theories also admit the accelerated expansion of the Universe without resorting to DE. It is remarkably that their equations of motion are always of second order in contrast with GR where the field equations are fourth order equations. Different cosmological properties of modified $F(T)$ gravity models  have been studied in the literature  (see e.g. \cite{FF1}-\cite{Bamba011}). Both possibilities have, in principle, the capability to explain the accelerated expansion of the Universe and even the primordial inflationary phase, with no need to introduce a new form of energy.

In this paper we will study some specific  $F(T)$ torsion gravity theories and, by using a technique developed in \cite{N4}, the corresponding cosmological theory will be reconstructed for several cosmological solutions. We perform a number of explicit reconstructions which lead to a number of interesting analytical results.

\section{The  model}
Let us we start with the following action for the $F(T)$ gravity \cite{FF1} - \cite{FF2}
\begin {equation}
S=\frac{1}{2k^2}\int dx^{4}[\sqrt{-g}F(T)+L_{m}],
\end{equation}
where $T$ is the torsion scalar, $F(T)$ is general differentiable function of the torsion and $L_{m}$ corresponds to the matter Lagrangian, $k^2=8\pi G$. Here the torsion scalar $T$ is defined as 
\begin{align}
	T=S_\rho^{\mu\nu}T^\rho_{\mu\nu},
\end{align}
where
\begin{align}
	S_\rho\,^{\mu\nu}=\frac{1}{2}(K^{\mu\nu}\,_\rho+\delta^\mu_\rho T^{\theta\nu}\,_\theta-\delta^\nu_\rho T^{\theta\mu}\,_\theta),
\end{align}
\begin{align}
K^{\mu\nu}\,_\rho=-\frac{1}{2}(T^{\mu\nu}\,_\rho-T^{\nu\mu}\,_\rho-T_\rho\,^{\mu\nu}),
\end{align}
\begin{align}
T^\lambda_{\mu\nu}=\stackrel{w}{\Gamma}^\lambda_{\nu\mu}-\stackrel{w}{\Gamma}^\lambda_{\mu\nu}=e^\lambda_i(\partial_\mu e^i_\nu-\partial_\nu e^i_\mu).
\end{align}
Here we have  the following formulas:
\begin{align}
g_{\mu\nu}(x)=\eta_{ij}e^i_\mu(x)e^j_\nu(x), 
\end{align}
where
\begin{align}
{\bf e}_i \cdot {\bf e}_j=\eta_{ij}, \quad  \eta_{ij}=diag(1, -1, -1, -1).
\end{align}
The field equations for the action (2.1) reads   \cite{FF1} - \cite{FF2}
\begin {equation}
[e^{-1}\partial_{\mu}(eS^{\mu\nu}_{i}-e^{\lambda}_{i}T^{\rho}_{\mu\lambda}S^{\nu\mu}_{\rho}]F_{T}+S^{\mu\nu}_{i}\partial_{\mu}TF_{TT}+\frac{1}{4}e^{\nu}_{i}F=\frac{1}{2}k^2e^{\rho}_{i}T^{\nu}_{\mu},
\end{equation}
where $e=\sqrt{-g}$. We now will assume a flat homogeneous and isotropic FRW universe with  the metric 
\begin {equation}
ds^{2}=-dt^{2}+a(t)^{2}\sum^{3}_{i=1}(dx^{i})^{2},
\end{equation}
where $t$ is  cosmic time. For this metric  
\begin{align}
	e_\mu^i=diag(1, a(t), a(t), a(t)), \quad H=\frac{\dot{a}}{a}, \quad T=-6H^2
\end{align}
and   the modified Friedmann equations and the conservation equation  reads as
\begin{align}
	12H^2 F_{T}+F=2k^2 \rho, 
\end{align}
\begin{align}
	48 H^2 \dot{H}F_{TT}-(12H^2+4\dot{H})F_{T}-F=2k^2p, 
\end{align}
\begin{align}
	\dot{\rho}+3H(\rho+p)=0.
\end{align}
This set of equations we can rewrite as
\begin{align}
	-2TF_{T}+F=2k^2 \rho, 
\end{align}
\begin{align}
	-8\dot{H}TF_{TT}+(2T-4\dot{H})F_{T}-F=2k^2p, 
\end{align}
\begin{align}
	\dot{\rho}+3H(\rho+p)=0.
\end{align}
 Note that the last equation can be written as  
\begin{align}
	\frac{d}{dt}(a^3\rho)=-3a^3Hp.
\end{align}
Also we note that as $F(T)=T$ Eqs. (2.14)-(2.15) transform to the usual Friedmann equations of GR
\begin{equation}
\frac{3}{k^2}H^2=\rho, \quad \frac{1}{k^2}(2\dot{H}+3H^2)=-p.
\end{equation}
We can rewrite  equations (2.14) - (2.15) as
\begin{equation}
\frac{3}{k^2}H^2=\rho+\rho_T, \quad \frac{1}{k^2}(2\dot{H}+3H^2)=-(p+p_T),
\end{equation}
where
\begin{equation}
\rho_T=\frac{1}{2k^2}(2TF_T-F+6H^2), \quad p_T=-\frac{1}{2k^2}[-8\dot{H}TF_{TT}+(2T-4\dot{H})F_{T}-F+4\dot{H}+6H^2]
\end{equation}
are the torsion contributions to the energy density and pressure. We also present the parameter of state 
\begin{align}
	w=-1+\frac{-8\dot{H}TF_{TT}-4\dot{H}F_{T}}{-2TF_{T}+F}.
\end{align}

\section{Reconstruction of a $F(T)$ theory: $\Lambda$CDM model}

Let us reconstruct the $F(T)$ theories starting from the $\Lambda$CDM model. As is known, in this case
\begin{align}
H^2=H_0^2+\frac{k^2\rho_0}{3a^3}=-\frac{1}{6}T
\end{align}
or
\begin{align}
T=-6H_0^2-\frac{2k^2\rho_0}{a^3},\quad a^{-1}=[\frac{3}{k^2\rho}(H^2-H_0^2)]^{\frac{1}{3}}=-[\frac{1}{2k^2\rho_0}(T+6H_0^2)]^{\frac{1}{3}}.
\end{align}
The solution to the homogeneous case of  equation (2.14) (that is, $\rho=0$)  is given by
\begin{align}
F=C\sqrt{T}, \quad C=const.
\end{align}
The second equation (2.15) satisfies automatically and gives $p=0$. Let us now explicitly reconstruct the $F(T)$ gravity theories for which a given matter field would obey a $\Lambda$CDM expansion history.
\subsection{Reconstruction for dust-like matter}
First we consider the case when the universe is filled with dust-like matter. Then from the energy conservation equation (2.16)  we have (as $w=0$)
\begin{align}
\rho=\rho_ca^{-3}=-\frac{\rho_c}{2k^2\rho_0}(T+6H_0^2).
\end{align}
Substituting in the modified Friedmann equation (2.14), for the unknown function $F(T)$, we get the  equation 
\begin{align}
	2TF_{T}-F=\frac{\rho_c}{\rho_0}(T+6H_0^2). 
\end{align}
The particular solution of this equation is
\begin{align}
	F(T)=\frac{\rho_c}{\rho_0}(T-6H_0^2).
\end{align}

\subsection{Reconstruction for perfect fluid with $w=-\frac{1}{3}$}

Let us now consider the perfect fluid with  $w=-\frac{1}{3}$.  Then   the energy conservation equation (2.16) gives
\begin{align}
\rho=\rho_ca^{-2}=\rho_c[\frac{1}{2k^2\rho_0}(T+6H_0^2)]^{\frac{2}{3}},\quad \rho_c=const.
\end{align}
The modified Friedmann equation (2.14) looks like
\begin{align}
	-2TF_{T}+F=2k^2\rho_c[\frac{1}{2k^2\rho_0}(T+6H_0^2)]^{\frac{2}{3}} =\alpha(T+\beta)^{\frac{2}{3}}. 
\end{align}
The  solution of this equation is given by
\begin{align}
F(T)=-\alpha\{-(T+\beta)^{\frac{2}{3}}+\frac{4}{3}\beta^{-\frac{1}{3}}THypergeometric2F1[\frac{1}{2}, \frac{1}{3},\frac{3}{2}, -\frac{x}{\beta}]\}.
\end{align}

\subsection{Reconstruction for stiff  fluid }

In this case from the energy conservation equation (2.16) we get
\begin{align}
\rho=\rho_sa^{-6}=\rho_c[\frac{1}{2k^2\rho_0}(T+6H_0^2)]^{2},\quad \rho_s=const.
\end{align}
In this case  the first modified Friedmann equation (2.14) reads 
\begin{align}
	-2TF_{T}+F=2k^2\rho_s[\frac{1}{2k^2\rho_0}(T+6H_0^2)]^{2} =\alpha(T+\beta)^{2}, 
\end{align}
which has the    solution 
\begin{align}
F(T)=-\frac{\alpha}{3}(T^2+6\beta T-3\beta^2).
\end{align}

\subsection{Reconstruction for two-fluids }

i) Now let us consider the more complicated case when the energy density has the form
\begin{align}
\rho=\rho_ca^{-2}+\rho_sa^{-6}=\rho_c[\frac{1}{2k^2\rho_0}(T+6H_0^2)]^{\frac{2}{3}}+\rho_s[\frac{1}{2k^2\rho_0}(T+6H_0^2)]^{2},
\end{align}
so that the first modified Friedmann equation (2.14) takes the form
\begin{align}
	-2TF_{T}+F=2k^2\{\rho_c[\frac{1}{2k^2\rho_0}(T+6H_0^2)]^{\frac{2}{3}}+\rho_s[\frac{1}{2k^2\rho_0}(T+6H_0^2)]^{2}\}
	 =\alpha(T+\beta)^{\frac{2}{3}}+\delta(T+\beta)^{2}.
\end{align}
The  solution of this equation is given by
\begin{align}
F(T)=-\frac{1}{3}[\delta (T^2+6\beta T-3\beta^2)-3\alpha(T+\beta)^{\frac{2}{3}}+4\alpha\beta^{-\frac{1}{3}}TZ],
\end{align}
where $Z=Hypergeometric2F1[\frac{1}{2}, \frac{1}{3},\frac{3}{2}, -\frac{x}{\beta}]$.

ii) Now we  consider the following example of two-fluid model
\begin{align}
\rho=\rho_ca^{-3}+\rho_sa^{-6}=\rho_c\frac{1}{2k^2\rho_0}(T+6H_0^2)+\rho_s[\frac{1}{2k^2\rho_0}(T+6H_0^2)]^{2}= \alpha(T+\beta)+\delta(T+\beta)^{2}.
\end{align}

The  corresponding solution of the modified Friedmann  equation reads 
\begin{align}
F(T)=-\frac{1}{3}[\delta  T^2+3(2\delta\beta +\alpha)T-3\delta\beta^2-3\alpha\beta].
\end{align}

iii) At last, we  consider the  example 
\begin{align}
\rho=\rho_ca^{-3}+\rho_sa^{-4}=\rho_c\frac{1}{2k^2\rho_0}(T+6H_0^2)+\rho_s[\frac{1}{2k^2\rho_0}(T+6H_0^2)]^{\frac{4}{3}}.
\end{align}
So  the first modified Friedmann equation (2.14) takes the form
\begin{align}
	-2TF_{T}+F=\rho_c\frac{1}{2k^2\rho_0}(T+6H_0^2)+\rho_s[\frac{1}{2k^2\rho_0}(T+
	6H_0^2)]^{\frac{4}{3}}=\alpha(T+\beta)+\delta(T+\beta)^{\frac{4}{3}},
\end{align}
which has the  following solution: 
\begin{align}
F(T)=-\frac{1}{15}[15\alpha(T-\beta)
-3\delta(5\beta-3T)(T+\beta)^{\frac{1}{3}}+16\delta\beta^{\frac{1}{3}} TY].
\end{align}
Here $Y=Hypergeometric2F1[\frac{1}{2},\frac{2}{3},\frac{3}{2},-\frac{x}{\beta}].$
\section{Reconstruction using $w$}
Now we will come to the reconstruction  problem from a slightly different position, namely, we reconstruct $F(T)$ for particular values of the EoS parameter  $w$. To do it, let us  consider  the  equation
\begin{align}
	-8\dot{H}TF_{TT}+[(1+w)2T-4\dot{H}]F_{T}-(1+w)F=0
\end{align}
that follows from (2.21). Consider examples.
\subsection{$w=-1$}
In this case  equation (4.1) takes the form
\begin{align}
\dot{H}(2TF_{TT}+F_{T})=0
\end{align}
This equation admits two solutions. i) $\dot{H}=0$. Hence $H=H_0=const$. It is the de Sitter spacetime. ii) The second solution obeys the equation
\begin{align}
	2TF_{TT}+F_T=0 
\end{align}
which has the solution
\begin{align}
	F(T)=2C_1T^{\frac{1}{2}}+C_2, \quad C_j=consts.
\end{align}
\subsection{$w=-1/3$}
In this case, Eq.(4.1) transforms to the form
\begin{align}
	-4\dot{H}TF_{TT}+2[\frac{1}{3}T-\dot{H}]F_{T}-\frac{1}{3}F=0.
\end{align}
Let $a=a_0(t-t_0)^n.$ Then Eq.(4.5) has the particular solution of the form  $F=T^n$. 
\subsection{$w=0$}
For this example, the pressure is equal to zero and Eq.(4.1) takes the form 
\begin{align}
	-8\dot{H}TF_{TT}+2[T-2\dot{H}]F_{T}-F=0
\end{align}
Let $a=a_0(t-t_0)^m$. Then Eq.(4.6) has the  solution  $F=T^{\frac{3m}{2}}$.

\section{$F(T)$ is given}

In this section, we assume that $F(T)$ is known (given) and find the corresponding expressions for $\rho$ and $ p$.

\subsection{$F=\alpha T+\beta T^{\delta}\ln{T}$}
As the first example we consider the following $F(T)$ model:
\begin{align}
F=\alpha T+\beta T^{\delta}\ln{T}.
\end{align}
Then the modified Friedmann equations give
\begin{align}
		\alpha T+2\beta T^{\delta}+\beta (2\delta-1)T^{\delta}\ln{T}=-2k^2 \rho,
\end{align}
\begin{align}
	\beta(2\delta-1) T^{\delta-1}(T-4\delta \dot{H})\ln{T}-4\beta(4\delta-1)\dot{H}T^{\delta-1}+2\beta T^\delta+\alpha T-4\alpha\dot{H}=2k^2p.
\end{align}
If $\delta=0.5$ then we get the more simple expressions for the energy density and pressure
\begin{align}
		\alpha T+2\beta T^{0.5}=-2k^2 \rho,
\end{align}
\begin{align}
	-4\beta\dot{H}T^{-0.5}+2\beta T^{0.5}+\alpha T-4\alpha\dot{H}=2k^2p,
\end{align}
respectively. If we also assume that  $\alpha=0$, then these formulas become
\begin{align}
		2\beta T^{0.5}=-2k^2 \rho,
\end{align}
\begin{align}
	-4\beta\dot{H}T^{-0.5}+2\beta T^{0.5}=2k^2p.
\end{align}

\subsection{$F=\alpha T+\frac{\beta}{ T}$}
Our  next example has the form
\begin{align}
	F=\alpha T+\frac{\beta}{ T}.
\end{align}
For this case we get
\begin{align}
	2k^2\rho=-\alpha T+\frac{3\beta}{T},
\end{align}
\begin{align}
	2k^2p= -12\beta\dot{H}T^{-2}-3\beta T^{-1}+
	 \alpha T-4\alpha\dot{H}.
\end{align}
The corresponding parameter of state reads 
\begin{align}
	w=-1+\frac{4\dot{H}(3\beta+\alpha T^2)}{T(\alpha T-3\beta)}.
\end{align}

\subsection{$F=\alpha T+\beta T^n$}
Finally, let us consider the pedagogical example 
\begin{align}
	F=\alpha T+\beta T^n.
\end{align}
We have
\begin{align}
	2k^2\rho=-\alpha T+\beta(1-2n)T^n,
\end{align}
\begin{align}
	2k^2p= \alpha T-\beta(1-2n)T^n+\beta(1-n)T^n-4\beta n(2n-1)\dot{H}T^{n-1}-4\alpha\dot{H}
\end{align}
and
\begin{align}
	w=-1-\frac{[\beta(1-n)T^n-4\beta n(2n-1)\dot{H}T^{n-1}-4\alpha\dot{H}]}{\alpha T-\beta(1-2n) T^n}.
\end{align}
In the case $n=2$ these expressions take more simple form as
\begin{align}
	2k^2\rho=-\alpha T-3\beta  T^2,
\end{align}
\begin{align}
	2k^2p= -24\beta\dot{H}T +3\beta T^2 +\alpha T-4\alpha\dot{H},
\end{align}
\begin{align}
	w=-1+\frac{4\dot{H}(6\beta T+\alpha)}{T(\alpha +3\beta T)}.
\end{align}

\section{Cosmic acceleration}
Let us we now consider the model
\begin{align}
p=\frac{A_{-1}(T)}{\rho}+A_0(T)+A_1(T)\rho.
\end{align}
Note that if $A_{-1}=const.,\quad  A_0=A_1=0$, it turns to the Chaplygin gas. We study  (6.1) for $F=\alpha T+\beta T^n$. We have
\begin{align}
	2k^2\rho=-\alpha T+\beta(1-2n)T^n
\end{align}
\begin{align}
	2k^2p= WT^{n-1}+V,
	\end{align}
	where
	\begin{align}
	W=	\beta(2n-1)(T-4 n\dot{H}), \quad V=\alpha(T-4\dot{H}).
\end{align}
Now for simplicity we assume that $n=0.5, \quad A_k=consts$ and $2k^2=1$. Then
\begin{align}
\dot{H}=\frac{\alpha^2(A_1+1)T^2+A_{-1}-\alpha A_0 T}{4\alpha^2 T}
\end{align}
or
\begin{align}
T_{N}=-3[\frac{\alpha^2(A_1+1)T^2+A_{-1}-\alpha A_0 T}{\alpha^2 T}]=xT+y+\frac{z}{T},\quad N=\ln{\frac{a}{a_0}}.
\end{align}
This equation has the  solution
\begin{align}
(2x)^{-1}\{-2y(4xz-y^2)^{-\frac{1}{2}}ArcTan{[(y+2xT)(4xz-y^2)^{-\frac{1}{2}}]}+
Log{[z+T(y+xT)]}\}=N-N_0.
\end{align}
Hence, for instance, for the  case $A_0=A_1+1=0$, we get the following solution
\begin{align}
T=[2z(N-N_0)]^{\frac{1}{2}}
\end{align}
or
\begin{align}
H=\{-\frac{1}{6}[2z(N-N_0)]^{\frac{1}{2}}\}^{\frac{1}{2}}.
\end{align}

Now let us consider the power-law solution for the scale factor:  $	a=a_0t^m$. Then for the model (6.8) - (6.9) we get the expression for the EoS parameter:
\begin{align}
	w=-1+4\frac{\dot{H}}{T}=-1-\frac{T_N}{3T}=
	-1-\frac{z}{3T^2}=-1-\frac{1}{6(N-N_0)}.
\end{align}
So  we can conclude that some of the above presented  models admit  cosmic acceleration.
Finally, we present some  examples of Weierstrass cosmologies. For example,  the Weierstrass gas which  has the following EoS
 \begin{equation}
 p=-B[\wp(\rho)]^{0.5}.\end{equation}
 The well-known Chaplygin gas model 
  \begin{equation}
 p=-\frac{B}{\rho}\end{equation}
 is the limit (or particular case) of the Weierstrass gas when the Weierstrass function takes the form
  \begin{equation}
 \wp(\rho)=\rho^{-2}.\end{equation}
  The generalized Weierstrass gas corresponds to the  EoS
  \begin{equation}
 p=-B[\wp(\rho)]^{0.5\alpha}.\end{equation}
 Its limit (or the particular case) is the generalized Chaplygin gas   \begin{equation}
 p=-\frac{B}{\rho^{\alpha}}.\end{equation}
 The modified Weierstrass gas (MWG) model  is given by
   \begin{equation}
 p=A\rho-B[\wp(\rho)]^{0.5\alpha}.\end{equation}
 The MWG is some generalization of the modified Chaplygin gas 
   \begin{equation}
 p=A\rho-\frac{B}{\rho^{\alpha}}.\end{equation}
 Finally we would like to give the more general form of the MWG. Its EoS reads as
  \begin{equation}
 p=A[\wp(\rho)]^{-0.5}-B[\wp(\rho)]^{0.5\alpha}.\end{equation}
 
\section{Conclusion}
In the present paper, several types of DE cosmology in $F(T)$  gravity have been investigated. We have studied different kinds of theory, in all of which the torsion invariant plays an important role in the corresponding equations. We have shown that the modified torsion gravity
 provides a very powerful theory, where no sort of DE is actually needed to reproduce the standard $\Lambda$CDM cosmology.

So we can conclude that $F(T)$  gravity theory studied in the paper adequately reproduces any kind of cosmological solution. As a follow up, we have shown, with the help of several particular examples corresponding to explicit choices of the function $F(T)$,  in principle, any cosmic evolution can be obtained from these models, what includes e.g. the unification of early-time inflation with the late-time acceleration coming from astronomical observations.  Also it is important to study the relationship $F(T)$ gravity with k-essence, f-essence and g-essence (see e.g.  \cite{MR23}, \cite{MR24} and  \cite{MR25} - \cite{MR29}). Finally we would like to note that one of interesting  classes of $F(T)$ gravity is integrable models. These models can be constructed by using the methods of  \cite{MR30} - \cite{MR32}. These important problems will be subjects of the future investigations. 
-

\end{document}